## Short Paper

Collin Sakal, MSc[1]; Tingyou Li[1]; Juan Li, PhD[2]; Xinyue Li, PhD[1]*

Affiliations

1. School of Data Science, City University of Hong Kong, Hong Kong SAR, China
2. Center on Aging Psychology, Key Laboratory of Mental Health, Institute of Psychology, Chinese Academy of Sciences, Beijing, China

*Address correspondence to: Xinyue Li, PhD
Xinyue Li, PhD
School of Data Science, City University of Hong Kong, Hong Kong SAR, China
Postal address: 83 Tat Chee Avenue, Lau-16-224, Kowloon Tong, Hong Kong SAR, China
Telephone number: (+852)34422180
Email: xinyueli@cityu.edu.hk

# Associations Between Sleep Efficiency Variability and Cognition Among Older Adults: Cross-Sectional Accelerometer Study

## Abstract

**Background:**
Sleep efficiency is often used as a measure of sleep quality. Getting sufficiently high-quality sleep has been associated with better cognitive function among older adults, however, the relationship between day-to-day sleep quality variability and cognition has not been well-established.

**Objective:**
We aimed to determine the relationship between day-to-day sleep efficiency variability and cognitive function among older adults using accelerometer data and three cognitive tests.

**Methods:**
Older adults aged 65+ with at least 5 days of accelerometer wear time from the National Health and Nutrition Examination Survey (NHANES) who completed the Digit Symbol Substitution Test (DSST), the Consortium to Establish a Registry for Alzheimer's Disease Word-Learning subtest (CERAD-WL), and Animal Fluency Test (AFT) were included in this study. Sleep efficiency was derived using a data-driven machine learning algorithm. Associations between sleep efficiency variability and scores on each cognitive test were examined adjusted for age, sex, education, household income, marital status, depressive symptoms, diabetes, smoking habits, alcohol consumption, arthritis, heart disease, prior heart attack, prior stroke, activities of daily living, and instrumental activities of daily living. Associations between average sleep efficiency and each cognitive test were further examined for comparison purposes.

**Results:**
A total of 1074 older adults from the NHANES were included in this study. Older adults with low average sleep efficiency exhibited higher levels of sleep efficiency variability (Pearson's r = -0.63). After adjusting for confounding factors, greater average sleep efficiency was associated with higher scores on the DSST (per 10% increase, β 2.25, 95% CI 0.61 to 3.90) and AFT (per 10% increase, β 0.91, 95% CI 0.27 to 1.56). Greater sleep efficiency variability was univariably associated with worse cognitive function based on the DSST (per 10% increase, β -3.34, 95% CI -5.33 to -1.34), CERAD-WL (per 10% increase, β -1.00, 95% CI -1.79 to -0.21), and AFT (per 10% increase, β -1.02, 95% CI -1.68 to -0.36). In the fully adjusted models, greater sleep efficiency variability remained associated with lower DSST (per 10% increase, β -2.01, 95% CI -3.62 to -0.40) and AFT (per 10% increase, β -0.84, 95% CI -1.47 to -0.21) scores but not CERAD-WL scores.

**Conclusions:**

Targeting consistency regarding sleep quality may be useful for interventions seeking to preserve cognitive function among older adults.

**Keywords:** aging; cognition; accelerometer; sleep; sleep efficiency; geriatrics; gerontology; actigraphy; digital health

## Introduction

Healthy sleep habits are protective of memory and cognitive function [1], but our understanding of the associative factors between sleep and cognition is incomplete. Sleep quality deteriorates with age, but older adults with cognitive impairments are known to have worse sleep quality than their unimpaired counterparts [2, 3]. Lower sleep efficiency, a proxy for sleep quality, is associated with worse cognition among older adults [4]. The importance of getting sufficiently high-quality sleep to reduce individual risk of cognitive impairments has been established in existing studies [5]. However, the role of consistency regarding sleep quality and its relationship with cognition remains understudied. Given that it is unreasonable to assume older adults strictly adhere to a consistent sleep schedule on a nightly basis, the relationship between day-to-day sleep efficiency variability and cognition must be examined.

In this cross-sectional accelerometer study we aimed to quantify associations between sleep efficiency variability with three cognitive tests that assess memory, categorical verbal fluency, and sustained attention. Associations were examined after adjusting for demographic factors, chronic diseases, smoking habits, alcohol consumption, cardiovascular risk factors, depressive symptoms, and measures of activities of daily living (ADL) and instrumental activities of daily living (IADL). We additionally fit models using average sleep efficiency metrics to compare any observed relationships between sleep efficiency variability and cognition to those between average sleep efficiency and cognition.

## Methods

### Data Source and Study Design

The data used in this study was taken from the 2011-2014 waves of the National Health and Nutrition Examination Survey (NHANES) [6] in the United States. During the 2011-14 NHANES waves a subset of participants wore an ActiGraph GT3X+ that objectively measured activity levels over seven consecutive days. Participants over the age of 60 were also administered cognitive tests during the 2011-14 waves. All NHANES participants gave their informed consent and ethics approval was granted by the National Center for Health Statistics Research Ethics Review Board. We excluded NHANES participants who were less than 65 years old, without complete cognitive test data, or did not have at least five days of accelerometer wear time.

### Measuring Cognition

The NHANES 2011-2014 waves include three cognitive tests: the Digit Symbol Substitution Test (DSST), the Consortium to Establish a Registry for Alzheimer's Disease Word-Learning subtest (CERAD-WL), and Animal Fluency Test (AFT). The AFT measures verbal categorical fluency and requires participants to name as many animals as possible in a one-minute period. The CERAD-WL measures immediate and delayed word recall. Three rounds of immediate recall and one round of delayed recall using lists of ten unrelated words comprise the CERAD-WL. Scores for the

CERAD-WL correspond to the number of correctly recalled words across all three rounds. The DSST tests processing speed, sustained attention, and working memory. For the DSST each participant was first given a key mapping nine numbers to nine symbols before being tasked with writing as many symbols as possible below 133 numbers in a two-minute period with possible scores ranging from 0-133. For the AFT, CERAD-WL, and DSST higher scores correspond to better cognition.

### Deriving Sleep Metrics

Sleep efficiency is a proxy for sleep quality and represents the ratio of time spent asleep over time dedicated to sleep. Possibly values for sleep efficiency range from zero to one with higher values corresponding to better quality sleep. Nightly sleep efficiency values were derived using an unsupervised Hidden Markov Model (HMM) that identifies sleep-wake states using a data-driven machine learning approach [7]. Sleep efficiency variability was defined as the standard deviation of sleep efficiency across all nights of valid accelerometer data. For comparison purposes we further derived each participant's average sleep efficiency.

### Additional Covariates

To account for potential confounding, we gathered each participant's age, sex, education, marital status, household income, smoking habits, current alcohol consumption, depressive symptoms, measures of functional independence, history of heart attack, history of stroke, and diagnoses of arthritis, heart disease, and diabetes. Depressive symptoms were defined using scores from the nine-item Patient Health Questionnaire (PHQ-9) [8]. A functional independence score was derived by summing responses to twenty activities of daily living and instrumental activities of daily living questions. Participants were categorized as current, former, or never smokers and drinkers. An explicit explanation of how each covariate was defined can be found in Multimedia Appendix 1. Participants with missing data were excluded to enable a complete-case analysis.

### Statistical Analysis

Participant characteristics were reported using the mean and standard deviation for numeric variables and counts and percentages for categorical variables. We first examined the relationship between mean and day-to-day sleep efficiency variability using Pearson's r correlation and a scatterplot. Thereafter, using cutoffs from previous studies [9], we plotted the distribution of sleep efficiency variability stratified by normal sleep efficiency (≥0.85) versus low sleep efficiency (<0.85).

We examined univariable associations between sleep efficiency variability with DSST, CERAD-WL, and AFT scores. Demographic models were then considered that adjusted for age, sex, education, marital status, and household income. Finally, the full models in this study further adjusted for depressive symptoms, ADL/IADL scores, smoking habits, alcohol consumption, diabetes, arthritis, heart disease, history of stroke, and history of heart attack. All univariable, demographic, and full models were re-fit using average sleep efficiency instead of day-to-day variability for comparison purposes. A sensitivity analysis was then conducted where we excluded extreme outliers (observations ≤1st quantile or ≥99th quantile) for both average sleep efficiency and day-to-day variability.

## Results

### Descriptive Statistics

A total of 1074 NHANES participants were included for the final analysis in this study (Table 1, Table S1 in Multimedia Appendix 1). The average age was 72.3 (5.2) and 49% of participants were female. The average sleep efficiency in the cohort was 0.94 (0.05), while the average DSST, CERAD-WL, and AFT scores were 46.7 (16.0), 25.0 (6.29), and 16.8 (5.25) respectively. The correlation between mean and day-to-day sleep efficiency variability was -0.63 (Figure 1). We found that older adults with low average sleep efficiency had higher levels of sleep efficiency variability compared to older adults with normal sleep efficiency levels (Figure 2).

**Table 1**. Demographic, sleep, and cognitive characteristics of NHANES older adults with valid accelerometer and cognitive test data

| Characteristic | Participants (n=1074) |
| --- | --- |
| Age (years), mean (SD) | 72.3 (5.2) |
| **Sex, n(%)** | |
|     Male | 546 (0.51) |
|     Female | 528 (0.49) |
| **Education, n(%)** | |
|     Less than 9th grade | 95 (0.09) |
|     Some high school | 141 (0.13) |
|     High school grad/GED | 245 (0.23) |
|     Some college or associates degree | 307 (0.29) |
|     College graduate or above | 286 (0.27) |
| **Marital Status, n(%)** | |
|     Married | 613 (0.57) |
|     Widowed | 230 (0.21) |
|     Divorced | 145 (0.14) |
|     Separated | 19 (0.02) |
|     Never married | 42 (0.04) |
|     Living with partner | 25 (0.02) |
| Sleep efficiency variability, mean (SD) | 0.04 (0.05) |
| Average sleep efficiency, mean (SD) | 0.94 (0.05) |
| DSST score, mean (SD) | 46.7 (16.0) |
| CERAD-WL score, mean (SD) | 25.0 (6.29) |
| AFT score, mean(SD) | 16.8 (5.25) |

**Figure 1.** Scatterplot with fitted regression line of average versus day-to-day variability for sleep efficiency

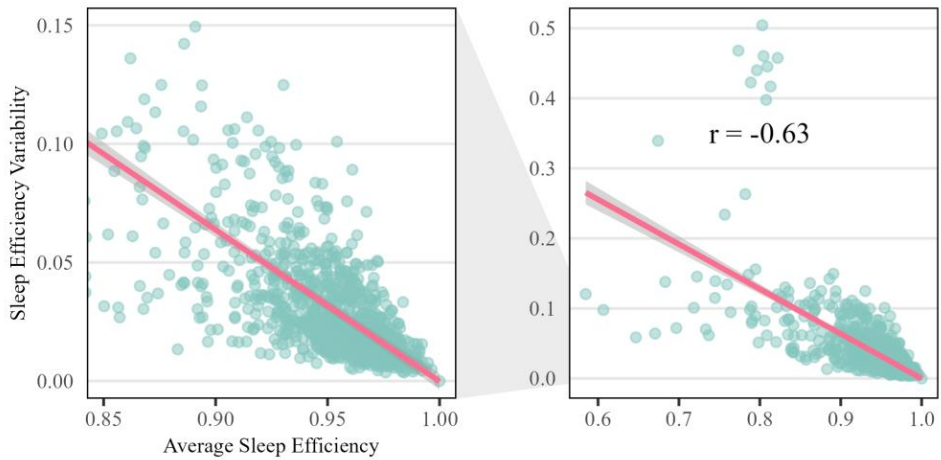

**Figure 2.** Distribution of sleep efficiency day-to-day variability stratified by average sleep efficiency

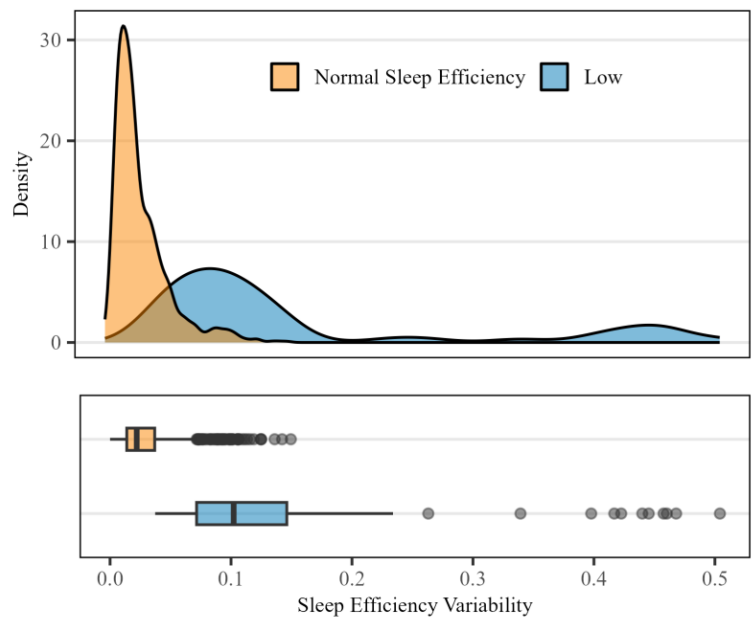

## Associations

In the univariable models greater sleep efficiency variability was associated with lower scores on the DSST (per 10% increase, β -3.34, 95% CI -5.33 to -1.34), CERAD-WL (per 10% increase, β -1.00, 95% CI -1.79 to -0.21), and AFT (per 10% increase, β -1.02, 95% CI -1.68 to -0.36). In the full models greater sleep efficiency variability was associated with lower scores on the DSST (per 10% increase, β -2.01, 95% CI -3.62 to -0.40) and the AFT (per 10% increase, β -0.84, 95% CI -1.47 to -0.21), but not the CERAD-WL (per 10% increase, β -0.65, 95% CI -1.39 to 0.08) (Tables 2-4). Conversely, greater average sleep efficiency was associated with higher scores on the DSST (per 10% increase, β 2.25, 95% CI 0.61 to 3.90) and AFT (per 10% increase, β 0.91, 95% CI 0.27 to 1.56) but not the CERAD-WL in the full models. In the sensitivity analysis, after excluding extreme average and sleep efficiency variability outliers, all significant results observed in the full models remained significant (Multimedia Appendix 2).

**Table 2.** Associations between day-to-day variability and average sleep efficiency with Digit Symbol Substitution Test (DSST) scores

|  | Association with DSST scores | |
|---|---|---|
| **Model Covariates** | β (95% CI)[a] | *P*-value |
| Sleep efficiency variability | -3.34 (-5.33, -1.34) | .001 |
| Mean sleep efficiency | 4.28 (2.27, 6.28) | <.001 |
| Demographics + sleep efficiency variability | -2.04 (-3.69, -0.39) | .016 |
| Demographics + mean sleep efficiency | 2.65 (0.97, 4.32) | .002 |
| Full model + sleep efficiency variability | -2.01 (-3.62, -0.40) | .015 |
| Full model + average sleep efficiency | 2.25 (0.61, 3.90) | .007 |

[a]Coefficients are reported per 10% increase

**Table 3.** Associations between day-to-day variability and average sleep efficiency with Consortium to Establish a Registry for Alzheimer's Disease Word-Learning subtest (CERAD-WL) scores

|  | Association with CERAD-WL scores | |
|---|---|---|
| **Model Covariates** | β (95% CI)[a] | *P*-value |
| Sleep efficiency variability | -1.00 (-1.79, -0.21) | .013 |
| Mean sleep efficiency | 0.85 (0.06, 1.65) | .036 |
| Demographics + sleep efficiency variability | -0.70 (-1.43, 0.03) | .061 |
| Demographics + mean sleep efficiency | 0.52 (-0.23, 1.26) | .175 |
| Full model + sleep efficiency variability | -0.65 (-1.39, 0.08) | .08 |
| Full model + average sleep efficiency | 0.46 (-0.29, 1.21) | .229 |

[a]Coefficients are reported per 10% increase

**Table 4.** Associations between day-to-day variability and average sleep efficiency with Animal Fluency Test (AFT) scores

|  | Association with AFT scores | |
|---|---|---|
| **Model Covariates** | β (95% CI)[a] | *P*-value |
| Sleep efficiency variability | -1.02 (-1.68, -0.36) | .002 |
| Mean sleep efficiency | 1.08 (0.42, 1.74) | .001 |
| Demographics + sleep efficiency variability | -0.85 (-1.48, -0.22) | .009 |
| Demographics + mean sleep efficiency | 1.02 (0.38, 1.66) | .002 |
| Full model + sleep efficiency variability | -0.84 (-1.47, -0.21) | .009 |
| Full model + average sleep efficiency | 0.91 (0.27, 1.56) | .005 |

[a]Coefficients are reported per 10% increase

## Discussion

### Principal Results and Comparisons with Prior Work

In this study we found that older adults with higher sleep efficiency variability scored worse on the Digit Symbol Substitution Test (DSST) and the Animal Fluency Test (AFT) after adjusting for demographic factors, chronic diseases, smoking habits, alcohol consumption, depressive symptoms, cardiovascular risk factors, and ADL/IADL scores. Effect sizes for average and sleep efficiency variability were similar in magnitude but in opposite directions, with greater variability being associated with lower DSST and AFT scores while greater average sleep efficiency was associated with higher scores.

One previous accelerometer study found that greater sleep efficiency variability was associated with lower scores on serial subtraction tests and memory questionnaires [10]. However, the study was limited by a small sample (N < 50) and did not consider relevant confounders such as chronic diseases, ADLs and IADLS, smoking habits, and alcohol consumption. Another study found that greater sleep efficiency variability was associated with greater β-amyloid burden, a biomarker for Alzheimer's disease, but was again limited by a small sample (N = 52) [11]. As such, compared to existing studies, our work provides evidence from a larger cohort accounting for more confounding factors that greater sleep efficiency variability is associated with worse cognitive function among older adults. Furthermore, we found that effect sizes for sleep efficiency variability and average sleep efficiency similar but in opposite directions, suggesting that getting sufficient and consistent high-quality sleep may be useful targets for interventions seeking to preserve cognitive function among older adults.

## Limitations

Given the cross-sectional nature of this study it is not clear if there is a causal relationship between sleep efficiency variability and poor cognition. Indeed, bi-directional associations are known to exist between certain forms of cognitive impairment and sleep disturbances [12]. As such, future studies may wish to examine if greater sleep efficiency variability is associated with future cognitive impairments in an unimpaired baseline cohort.

## Conclusions

Greater day-to-day sleep efficiency variability was associated with lower scores on two cognitive tests in the present study. While longitudinal studies are needed for confirmation, our work provides evidence that clinical recommendations aimed at protecting cognitive function among older adults may benefit from targeting consistency regarding sleep quality.

## Acknowledgements

CS, XL, and JL designed the study. CS and TL performed the statistical analyses with supervision from XL and JL. The manuscript was primarily written by CS with oversight from XL and JL. This work was supported by the City University of Hong Kong, Hong Kong SAR, China internal research grant #9610473. The funder played no role in study design, data collection, analysis and interpretation of data, or the writing of this manuscript.

## Conflicts of Interest

None declared

## Abbreviations

ADL: Activities of daily living
AFT: Animal fluency test

CERAD-WL: Consortium to establish a registry for Alzheimer's disease word-learning subtest
DSST: Digit symbol substitution test
IADL: Instrumental activities of daily living
NHANES: National Health and Nutrition Examination Survey
PHQ: Patient health questionnaire

## Multimedia Appendix 1
Complete cohort characteristics and variable derivation information

## Multimedia Appendix 2
Sensitivity analysis

## References

1. Scullin MK, Bliwise DL. Sleep, cognition, and normal aging: integrating a half century of multidisciplinary research. Perspect Psychol Sci. 2015 Jan;10(1):97-137. PMID: 25620997. doi: 10.1177/1745691614556680.
2. Livingston G, Huntley J, Sommerlad A, Ames D, Ballard C, Banerjee S, et al. Dementia prevention, intervention, and care: 2020 report of the Lancet Commission. The Lancet. 2020;396(10248):413-46. doi: 10.1016/s0140-6736(20)30367-6.
3. Unruh ML, Redline S, An MW, Buysse DJ, Nieto FJ, Yeh JL, et al. Subjective and objective sleep quality and aging in the sleep heart health study. J Am Geriatr Soc. 2008 Jul;56(7):1218-27. PMID: 18482295. doi: 10.1111/j.1532-5415.2008.01755.x.
4. D'Rozario AL, Chapman JL, Phillips CL, Palmer JR, Hoyos CM, Mowszowski L, et al. Objective measurement of sleep in mild cognitive impairment: A systematic review and meta-analysis. Sleep Med Rev. 2020 Aug;52:101308. PMID: 32302775. doi: 10.1016/j.smrv.2020.101308.
5. Mellow ML, Crozier AJ, Dumuid D, Wade AT, Goldsworthy MR, Dorrian J, et al. How are combinations of physical activity, sedentary behaviour and sleep related to cognitive function in older adults? A systematic review. Exp Gerontol. 2022 Mar;159:111698. PMID: 35026335. doi: 10.1016/j.exger.2022.111698.
6. Johnson CL, Dohrmann SM, Burt VL, Mohadjer LK. National health and nutrition examination survey: sample design, 2011-2014. Vital Health Stat 2. 2014 Mar(162):1-33. PMID: 25569458.
7. Li X, Zhang Y, Jiang F, Zhao H. A novel machine learning unsupervised algorithm for sleep/wake identification using actigraphy. Chronobiology International. 2020;37(7):1002-15. doi: 10.1080/07420528.2020.1754848.
8. Kroenke K, Spitzer RL, Williams JB. The PHQ-9: validity of a brief depression severity measure. J Gen Intern Med. 2001 Sep;16(9):606-13. PMID: 11556941. doi: 10.1046/j.1525-1497.2001.016009606.x.
9. Brindle RC, Yu L, Buysse DJ, Hall MH. Empirical derivation of cutoff values for the sleep health metric and its relationship to cardiometabolic morbidity: results from the Midlife in the United States (MIDUS) study. Sleep. 2019 Sep 6;42(9). PMID: 31083710. doi: 10.1093/sleep/zsz116.
10. Balouch S, Dijk DAD, Rusted J, Skene SS, Tabet N, Dijk DJ. Night-to-night variation in sleep associates with day-to-day variation in vigilance, cognition, memory, and behavioral problems in Alzheimer's disease. Alzheimers Dement (Amst). 2022;14(1):e12303. PMID: 35603140. doi: 10.1002/dad2.12303.

11. Fenton L, Isenberg AL, Aslanyan V, Albrecht D, Contreras JA, Stradford J, et al. Variability in objective sleep is associated with Alzheimer's pathology and cognition. Brain Commun. 2023;5(2):fcad031. PMID: 36895954. doi: 10.1093/braincomms/fcad031.

12. Ju YE, Lucey BP, Holtzman DM. Sleep and Alzheimer disease pathology--a bidirectional relationship. Nat Rev Neurol. 2014 Feb;10(2):115-9. PMID: 24366271. doi: 10.1038/nrneurol.2013.269.